\documentclass[preprint,12pt, nofootinbib]{revtex4}

\usepackage[brazil, english]{babel}
\usepackage[utf8]{inputenc}
\usepackage{graphicx}
\usepackage{epsfig}
\usepackage{amssymb}
\usepackage{amsmath} 
\usepackage{slashed}
\usepackage[unicode=true,bookmarks=false,breaklinks=false,pdfborder={0 0 1},colorlinks=true]
 {hyperref}
\hypersetup{
 citecolor=blue,linkcolor=blue,urlcolor=blue}

\graphicspath{%
    {converted_graphics/}
    {/}
}
\begin{document}

\title{AdS/BCFT correspondence and BTZ black hole thermodynamics within Horndeski gravity} 

\author{Fabiano F. Santos$^{1,2,}$}
\email[Eletronic address: ]{fabiano.ffs23@gmail.com}
\author{Eduardo Folco Capossoli$^{3,}$}
\email[Eletronic address: ]{eduardo\_capossoli@cp2.g12.br}
\author{Henrique Boschi-Filho$^{2,}$}
\email[Eletronic address: ]{boschi@if.ufrj.br}  
\affiliation{$^1$Instituto Federal de Educação, Ciência e Tecnologia do Sertão Pernambucano, 56316-686, Campus Petrolina Zona Rural, Pernambuco - PE,  Brazil\\$^2$Instituto de F\'{\i}sica, Universidade Federal do Rio de Janeiro, 21.941-972, Rio de Janeiro - RJ, Brazil\\
$^3$Departamento de F\'{\i}sica and Mestrado Profissional em Práticas de Educa\c{c}\~{a}o B\'{a}sica (MPPEB), 
 Col\'egio Pedro II, 20.921-903, Rio de Janeiro - RJ, Brazil}

\begin{abstract}
In this work, we study gravity duals of  conformal field  theories with boundaries$-$known as AdS/BCFT correspondence, put forward by Takayanagi  \cite{Takayanagi:2011zk}$-$within a scalar-tensor theory, as proposed by Horndeski  in 4D \cite{Horndeski}. We consider the case of 3D gravity dual to 2D BCFT, take a Gibbons-Hawking surface term modified by Horndeski's theory and find the corresponding 3D (Bañados-Teitelboim-Zanelli) black hole solutions. We analyze the effects of Horndeski gravity on the profile of the extra boundary for the  black hole by using an approximate analytic solution. Performing a holographic renormalization, we calculate the free energy and obtain the total entropy and corresponding area, as well as the boundary entropy for the  black hole. In particular, the boundary entropy found here can be seen as an extension of the one proposed by Takayanagi. From the free energy,  we perform a systematic study of the 3D black hole thermodynamics and present, among our results, an indication of the restoration of conformal symmetry for high temperatures. Finally, we present a study on the influence of the Horndeski gravity on the  Hawking-Page phase transition where we can see the stable and unstable phases throughout the plane of free energy versus temperature.
\end{abstract}


\maketitle


\section{Introduction}

Einstein's gravity is supported by both strong theoretical and experimental evidences. The observation of gravitational waves from a binary black hole merger, as reported in Ref. \cite{Abbott:2016blz}, was expected as one of the crucial tests for general relativity (GR). Despite this huge success, the last three decades brought some questions that could not be answered by GR. These questions are related to both theoretical aspects and observational results.  The very interesting review in Ref.  \cite{Capozziello:2011et} pointed out basically two classes of  ``shortcomings in GR'' on the  UV and IR scales. In the UV region one has the quantum gravity problem, and in the IR regime, the dark energy and dark matter issues. In order to address these  questions new approaches were proposed known as extended theories of gravity. Such theories start with the inclusion of higher-order terms in curvature invariants in the effective Lagrangian such as, for instance, $R^2$ and $R^{\alpha \beta \gamma \delta} R_{\alpha \beta \gamma \delta}$  \cite{Gottlober:1989ww, Adams:1990pn, Amendola:1993bg}, or through minimal or nonminimal coupling of scalar fields with the geometry such as, for example,  $\phi^2 R$  \cite{Maeda:1988ab, Wands:1993uu, Capozziello:1998dq}. 
The approach that takes into account a single scalar field in general relativity is known as Horndeski gravity \cite{Horndeski, Charmousis:2011bf, Charmousis:2011ea, Starobinsky:2016kua, Bruneton:2012zk, Cisterna:2014nua, Maselli:2016gxk,
 Heisenberg:2018vsk, Hajian:2020dcq}.
This model is quite interesting because it is the most general scalar-tensor theory with second-order field equations in four dimensions. 

Besides Horndeski gravity, in this work we consider two other essential components. The first one is related to the AdS/CFT correspondence or duality, whose fundamental concepts were explained in Refs. \cite{Maldacena:1997re, Gubser:1998bc, Witten:1998qj, Aharony:1999ti}. 
Over the past two decades since its emergence, many investigations around AdS/CFT have provided great insights on the study of strong coupled systems. Among many interesting features of this correspondence, one should note the possibility of building models on the gravity side that are duals to phases of a nonconformal plasma at finite temperature or density.  It is worthwhile to mention that in the recent Refs. \cite{Jiang:2017imk, Baggioli:2017ojd, Liu:2018hzo, Li:2018kqp, Li:2018rgn, Wang:2019jyw}, the authors presented some applications of AdS/CFT in the Horndeski scenario. 

The inclusion of another boundary in the original AdS/CFT duality leads to the AdS/BCFT correspondence, which has attracted a lot of attention in the last years. This proposal was presented by Takayanagi \cite{Takayanagi:2011zk} and soon after by Fujita, Takayanagi and Tonni \cite{Fujita:2011fp}  as an extension of the standard AdS/CFT correspondence.

The main point in AdS/CFT is based on the fact that the AdS$_{d+1}$ space is dual to a conformal field in $d$ dimensions. In this case the AdS$_{d+1}$ symmetry, which is $SO(2,d)$, is the same as the conformal symmetry of the CFT$_d$. However when one adds a new boundary with $d-1$ dimensions to the CFT$_d$, one notices the breaking of $SO(2,d)$  into a $SO(2,d-1)$ group. In this sense, due to the insertion of this new boundary, this theory is known as boundary conformal field theory (BCFT) and then one can construct a correspondence called AdS/BCFT  \cite{Takayanagi:2011zk, Fujita:2011fp, Nozaki:2012qd, Fujita:2012fp, Melnikov:2012tb, Magan:2014dwa, Erdmenger:2014xya, Erdmenger:2015spo, Flory:2017ftd}.\footnote{As pointed out in Ref. \cite{Fujita:2011fp} the relation between holography and BCFT was presented in the early 2000s as shown in Refs. \cite{Karch:2000ct, Karch:2000gx}. }  In particular, in this work we deal with an AdS$_3$/BCFT$_2$ correspondence. 

\begin{figure}[!ht]
\begin{center}
\includegraphics[scale=0.25]{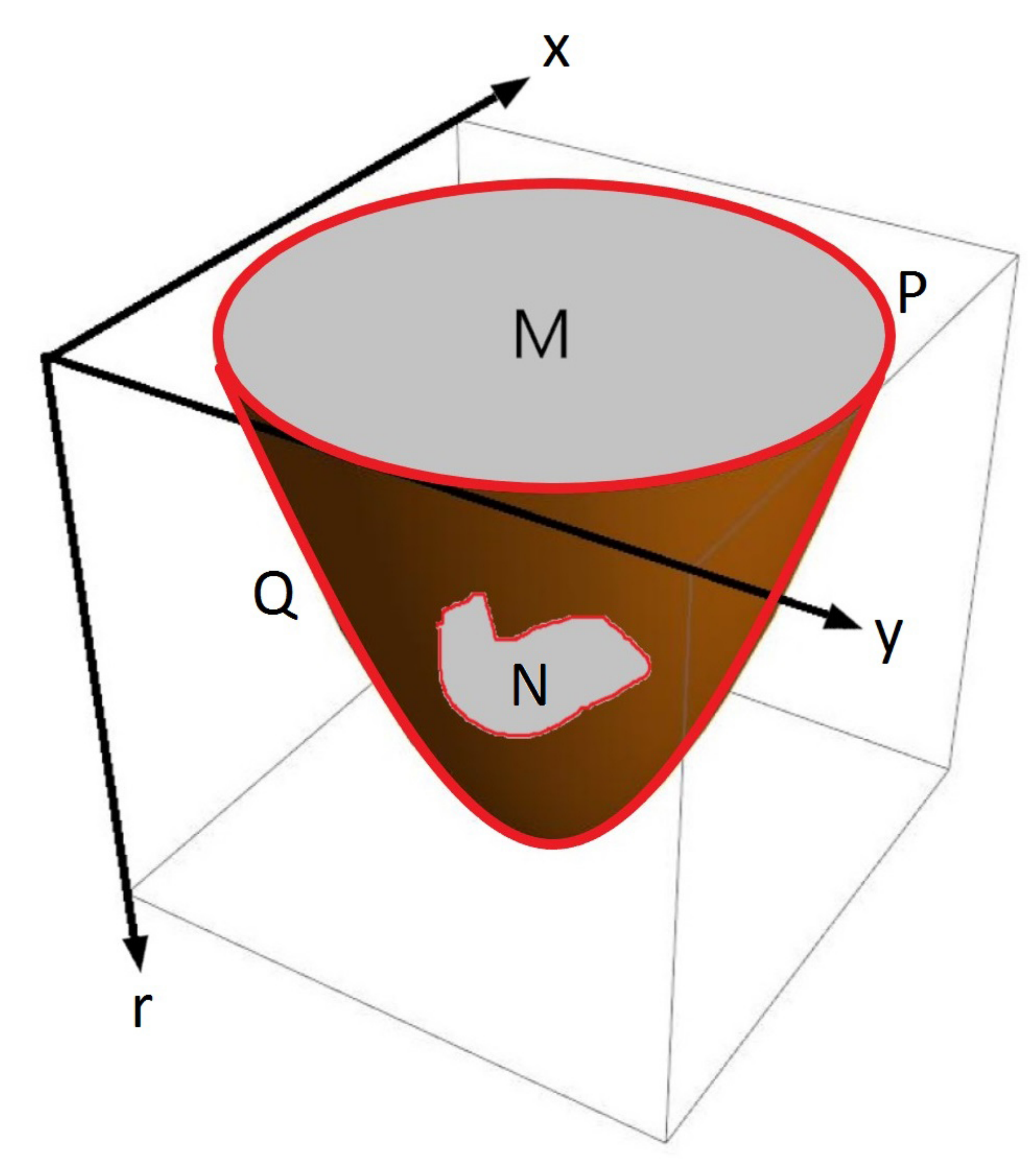}
\caption{Holographic description of BCFT, where we have the asymptotically AdS bulk spacetime N with conformal boundary M and additional boundary Q. Here P is the intersection of M and Q.}\label{BCFT}
\end{center}
\end{figure}

As we know, in the standard AdS/CFT correspondence we have an asymptotically anti–de Sitter (AdS) spacetime N, which has a boundary M with a Dirichlet boundary condition on it. On the other hand, following the AdS/BCFT prescription, we introduce an additional boundary\footnote{Note that the boundary Q in general is not asymptotically AdS.} Q wrapping N, whose interception with M is the manifold P, as shown in Fig. \ref{BCFT}. On the hypersurface Q, the bulk-metric of N should satisfy a Neumann boundary condition. Also by looking at Fig. \ref{BCFT} one should notice that the $d$-dimensional spacetime M is bounded by P, which will also bound Q. Within this construction, the $d+1$-dimensional spacetime N is limited by a region defined by $M \cup Q$.

%

The second essential component of this work is to deal with a finite-temperature theory within the AdS/CFT correspondence. Following the standard procedure we include a black hole in the bulk geometry and interpret the Hawking temperature as the temperature of the CFT side. 

In the past, $(2+1)$-dimensional gravity was considered as a toy model since (as pointed out in Ref. \cite{Carlip:1995qv}) it does not have a Newtonian limit nor any  propagating degrees of freedom. However, after the work of Bañados, Teitelboim, and Zanelli (BTZ) \cite{Banados:1992wn, Banados:1992gq}, it was realized that such a kind of (2+1) theory has a solution, known as the  BTZ black hole, with some interesting features: an event horizon (and in some cases an additional inner horizon, if one includes rotations) presenting thermodynamic properties somehow similar to the black holes in (3+1) dimensions and being asymptotically anti-de Sitter.\footnote{Usually, black holes are asymptotically flat.} 
For our purposes in this paper we choose to work with a planar BTZ black hole with a nontrivial axis profile.  

\section{Methodological route and achievements}

Motivated by the recent application of the AdS/CFT duality in Horndeski gravity together with the emergence of AdS/BCFT and taking into account the importance of (2+1)$-$dimensional black holes, in this work we establish the AdS/BCFT correspondence in Horndeski gravity and study the thermodynamics of the corresponding AdS-BTZ black hole. Here we present  a  summary  of  the  main  results  achieved  in  this  work:

\begin{itemize}

    \item First, we study the influence of the Horndeski parameters on the BCFT theory. Apart from a complete numerical solution, we derive an approximate analytical solution that is useful for determining the role of Q profile and perform an analysis of all quantities in this work; 
    
    \item We construct a holographic renormalization for this setup and compute the free energy for both the AdS-BTZ black hole and thermal AdS;
    
    \item From the free energy, we compute the total and boundary entropies. In the case of the boundary entropy, one can see it as an extension of the results found in Refs. \cite{Takayanagi:2011zk, Fujita:2011fp}; 
   
   \item Assuming that the total entropy and total area of the AdS-BTZ black hole are related by the Bekenstein-Hawking formula, we see that the influence of Horndeski gravity enables an increase of the black hole area as we increase the absolute value of the Horndeski parameter. This feature of our model is not present in  the usual BCFT theory as discussed, for instance, in Refs. \cite{Takayanagi:2011zk, Fujita:2011fp, Magan:2014dwa}. 
   
    \item  At zero temperature,  our setup exhibits a nonzero or residual boundary entropy, at least in certain conditions which depend on the tension of the Q profile. Besides, zero entropy seems to imply a minimum nonzero temperature. 
    
    \item From the free energy we also compute the thermodynamic observables, including the heat capacity, sound speed and trace anomaly and plot their behavior against the temperature. In particular, the trace anomaly goes to zero at high temperatures indicating a restoration of the conformal symmetry or a nontrivial  BCFT.  
    
    \item  We study the Hawking-Page phase transition (HPPT) in this setup. The presence of the Horndeski term allow us to analyze this transition of the free energy as a function of the temperature, as in other higher-dimensional theories. This differs from the results presented in Refs.  \cite{Takayanagi:2011zk, Fujita:2011fp}, where the authors plotted the free energy as a function of the tension of the Q profile. 
\end{itemize}

This work is organized as follows. In Section \ref{v1}, we present our gravitational setup and how to combine it with BCFT theory.  In the Section \ref{v2}, we consider a BTZ black hole in Horndeski gravity and study the influence of the Horndeski parameter on Q profile. In Section \ref{v3}, by performing a holographic renormalization  we compute the Euclidean on-shell actions associated with the BTZ and thermal AdS space. In Section \ref{BTZentro}, from the euclidean on-shell action, we derive the BTZ black hole entropy, and in Section \ref{v4}, we present a systematic study of its thermodynamic quantities. In section \ref{v5}, we present the Hawking-Page phase transition between the BTZ black hole and thermal AdS space. Finally, in Section \ref{v6} we present our conclusion and final comments.

\section{The Setup}\label{v1}
\subsection{Horndeski’s Lagrangian}

In this section, we present the outline of Horndeski gravity. The complete Horndeski Lagrangian density can be written in a general form as:
\begin{eqnarray}\label{LH}
{\cal L}_H = {\cal L}_{EH} + {\cal L}_2 + {\cal L}_3 + {\cal L}_4 + {\cal L}_5\,, 
\end{eqnarray}
\noindent where ${\cal L}_{EH}=\kappa(R-2\Lambda)$ is the Einstein-Hilbert Lagrangian density,   $R$ is the Ricci scalar, $\Lambda$ the cosmological constant, $\kappa=(16\pi G_N)^{-1}$, $G_N$ is Newton's gravitational constant,  and we choose $4 \pi G_N = 1$, so that $\kappa = 1/4$.  The Lagragians ${\cal L}_2$, ${\cal L}_3$, ${\cal L}_4$ and ${\cal L}_5$ are given by:\footnote{Since the publication of Ref. \cite{Charmousis:2011bf}, one usually refers to ${\cal L}_2, {\cal L}_3, {\cal L}_4$ and ${\cal L}_5$ in Eq. \eqref{4L} as the {\it Fab Four} Lagrangians.}
\begin{eqnarray}
{\cal L}_2 &=& G_2(X, \phi)\,, \nonumber \\ 
{\cal L}_3 &=& -G_3(X, \phi) \Box \phi\,, \nonumber \\ 
{\cal L}_4 &=& -G_4(X, \phi) R + \partial_X G_4(X, \phi) \delta^{\mu \nu}_{\alpha \beta} \nabla^{\alpha}_{\mu} \phi \nabla^{\beta}_{\nu} \phi\,, \nonumber \\ 
{\cal L}_5 &=& -G_5(X, \phi) G_{\mu \nu} \nabla^{\mu} \nabla^{\nu}\phi - \frac{1}{6} \partial_X G_5(X, \phi) \delta^{\mu \nu \rho}_{\alpha \beta \gamma} \nabla^{\alpha}_{\mu} \phi \nabla^{\beta}_{\nu} \phi \nabla^{\gamma}_{\rho} \phi \,, \label{4L}
\end{eqnarray}
\noindent 
 with $G_2$, $G_3$, $G_4$, and $G_5$ being arbitrary functions of the scalar field $\phi$ and $X$ is defined by $X \equiv - \frac{1}{2} \nabla_{\mu} \phi \nabla^{\mu} \phi$, while  $G_{\mu \nu}=R_{\mu\nu} -\frac 12 g_{\mu\nu}R$ is the Einstein tensor, and $g_{\mu\nu}$ is the spacetime metric. For a detailed review of Honderski gravity, one can see Ref. \cite{Kobayashi:2019hrl}. 

In particular, we are interested in a special subclass of Horndeski gravity that has a nonminimal coupling between the standard scalar term and the Einstein tensor
 \cite{Charmousis:2011bf,Charmousis:2011ea,Starobinsky:2016kua,Bruneton:2012zk,Brito:2019ose,Santos:2020xox}.  In this sense, Eq. \eqref{LH} becomes:
\begin{eqnarray}\label{HFE}
   {\cal L}_H \equiv {\cal L}_{EH} + {\cal L}_2 &=&(R-2\Lambda)-\frac{1}{2}(\alpha g_{\mu\nu}-\gamma\,  G_{\mu\nu})\nabla^{\mu}\phi\nabla^{\nu}\phi\,,
\end{eqnarray}
\noindent where $\alpha$ and $\gamma$ are the usual Horndeski  parameters, which control the strength of the kinetic couplings, and have mass dimensions 0 and $-2$, respectively. For convenience, we introduce the dimensionless parameter $\gamma'=-\gamma\Lambda$.  
Note that the Lagrangian density in Eq. \eqref{HFE},  is invariant under  displacement symmetry ($\phi\to\phi\, +\, constant$) and parity transformation ($\phi\to-\phi$).


\subsection{AdS$_3$/BCFT$_2$ correspondence with Horndeski $\gamma'$ dependence}

In this section, we discuss the AdS/BCFT correspondence within Horndeski gravity. 
 As discussed in Refs. \cite{Takayanagi:2011zk,Fujita:2011fp}, for the construction of boundary systems we need to take into account a Gibbons-Hawking surface term. In addition, such a surface term for Horndeski $\gamma'$-dependent gravity was proposed in Ref. \cite{Li:2018rgn}. Motivated by these works, we propose the total action including the contributions coming from the surfaces N, Q and P, besides matter terms from N and Q and the counterterms from P:\footnote{One can recall the AdS/BCFT geometry from Fig. \ref{BCFT}.}
 \begin{eqnarray}
S&=&S^{N}+S^{N}_{mat}+S^{Q}+S^{Q}_{mat}+S^{P}_{ct}\,, \label{S}
\end{eqnarray}
where $S^{N}_{mat}$ describes ordinary matter that is supposed
to be a perfect fluid, and 
\begin{eqnarray}
&&S^{N}=\kappa\int_{N}{d^{3}x\sqrt{-g}\mathcal{L}_{H}}\\
&&S^{Q}=2\kappa\int_{bdry}{d^{2}x\sqrt{-h}\mathcal{L}_{bdry}}\\ 
&&S^{Q}_{mat}=2\int_{Q}{d^{2}x\sqrt{-h}\mathcal{L}_{mat}}\\
&&S^{P}_{ct}=2\kappa\int_{ct}{d^{2}x\sqrt{-h}\mathcal{L}_{ct}}\,,
\end{eqnarray}
where $\mathcal{L}_{H}$ was defined in Eq. \eqref{HFE}  and 
\begin{eqnarray}
\mathcal{L}_{bdry}&=&(K-\Sigma)+\frac{\gamma'}{4\Lambda}(\nabla_{\mu}\phi\nabla_{\nu}\phi n^{\mu}n^{\nu}-(\nabla\phi)^{2})K+\frac{\gamma'}{4\Lambda}\nabla_{\mu}\phi\nabla_{\nu}\phi K^{\mu\nu}\,, \label{3}\\
\mathcal{L}_{ct}&=&c_{0}+c_{1}R+c_{2}R^{ij}R_{ij}+c_{3}R^{2}+b_{1}(\partial_{i}\phi\partial^{i}\phi)^{2}+...\label{4}
\end{eqnarray}
 Note that $\mathcal{L}_{mat}$ is a Lagrangian of possible matter fields on Q and $\mathcal{L}_{bdry}$ corresponds to the Gibbons-Hawking $\gamma'$-dependent terms associated with the Horndeski gravity. In the boundary Lagrangian, Eq. \eqref{3},  $K_{\mu\nu}=h^{\beta}_{\mu}\nabla_{\beta}n_{\nu}$ is the extrinsic curvature,
$h_{\mu\nu}$ is the induced metric and $n^\mu$ is the normal vector of the hypersurface Q. The traceless contraction of $K_{\mu\nu}$ is $K=h^{\mu\nu}K_{\mu\nu}$, and $\Sigma$ is the boundary tension on Q. Furthermore, ${\cal L}_{ct}$ are boundary counterterms localized on P which is required to be an asymptotic AdS spacetime.  By imposing a Neumann boundary condition in Eq. \eqref{3}, we obtain\footnote{For more details on the geometry, see Refs. \cite{Takayanagi:2011zk,Fujita:2011fp,Melnikov:2012tb,Magan:2014dwa}. Regarding the choice for the boundary condition, we refer to Ref. \cite{Compere:2008us} where the authors discussed a Neumann boundary condition, among others.}
\begin{eqnarray}
K_{\alpha\beta}-h_{\alpha\beta}(K-\Sigma)+\frac{\gamma'}{4\Lambda}H_{\alpha\beta}=\kappa {\cal S}^{Q}_{\alpha\beta}\,,\label{5}
\end{eqnarray}
where we defined 
\begin{eqnarray}
&&H_{\alpha\beta}\equiv(\nabla_{\alpha}\phi\nabla_{\beta}\phi n^{\alpha}n^{\beta}-(\nabla\phi)^{2})(K_{\alpha\beta}-h_{\alpha\beta}K)-(\nabla_{\alpha}\phi\nabla_{\beta}\phi)h_{\alpha\beta}K\,,\label{6}\\
&&{\cal S}^{Q}_{\alpha\beta}=-\frac{2}{\sqrt{-h}}\frac{\delta S^{Q}_{mat}}{\delta h^{\alpha\beta}}\,.\label{7} 
\end{eqnarray}
Considering $S^{Q}_{mat}$ as a constant one has ${\cal S}^{Q}_{\alpha\beta}=0$. Then, we can write
\begin{eqnarray}
K_{\alpha\beta}-h_{\alpha\beta}(K-\Sigma)+\frac{\gamma'}{4\Lambda}H_{\alpha\beta}=0\,.\label{8}
\end{eqnarray}
On the gravitational side, for Einstein-Horndeski gravity assuming that $S^{N}_{mat}$ is constant, and varying $S^N$ with respect to $g_{\alpha\beta}$ and $\phi$, and $S^Q$ with respect to $\phi$, respectively, we have:
\begin{eqnarray}
{\cal E}_{\alpha\beta}[g_{\mu\nu},\phi]=-\frac{2}{\sqrt{-g}}\frac{\delta S^{N}}{\delta g^{\alpha\beta}}\,,\quad {\cal E}_{\phi}[g_{\mu\nu},\phi]=-\frac{2}{\sqrt{-g}}\frac{\delta S^{N}}{\delta\phi} \,,\quad {\cal F}_{\phi}[g_{\mu\nu},\phi]=-\frac{2}{\sqrt{-h}}\frac{\delta S^{Q}}{\delta\phi} \,.\nonumber\\
\end{eqnarray}
Then, one finds: 
\begin{eqnarray}
{\cal E}_{\mu\nu}[g_{\mu\nu},\phi]&=&G_{\mu\nu}+\Lambda g_{\mu\nu}-\frac{\alpha}{2}\left(\nabla_{\mu}\phi\nabla_{\nu}\phi-\frac{1}{2}g_{\mu\nu}\nabla_{\lambda}\phi\nabla^{\lambda}\phi\right)\label{11}\nonumber\\
                  &-&\frac{\gamma'}{2\Lambda}\left(\frac{1}{2}\nabla_{\mu}\phi\nabla_{\nu}\phi R-2\nabla_{\lambda}\phi\nabla_{(\mu}\phi R^{\lambda}_{\nu)}-\nabla^{\lambda}\phi\nabla^{\rho}\phi R_{\mu\lambda\nu\rho}\right)\nonumber\\
									&-&\frac{\gamma'}{2\Lambda}\left(-(\nabla_{\mu}\nabla^{\lambda}\phi)(\nabla_{\nu}\nabla_{\lambda}\phi)+(\nabla_{\mu}\nabla_{\nu}\phi)\Box\phi+\frac{1}{2}G_{\mu\nu}(\nabla\phi)^{2}\right)\nonumber\\
									&+&\frac{\gamma' g_{\mu\nu}}{2\Lambda}\left(-\frac{1}{2}(\nabla^{\lambda}\nabla^{\rho}\phi)(\nabla_{\lambda}\nabla_{\rho}\phi)+\frac{1}{2}(\Box\phi)^{2}-(\nabla_{\lambda}\phi\nabla_{\rho}\phi)R^{\lambda\rho}\right),\\
{\cal E}_{\phi}[g_{\mu\nu},\phi]&=&\nabla_{\mu}\left[\left(\alpha g^{\mu\nu}+\frac{\gamma'}{\Lambda} G^{\mu\nu}\right)\nabla_{\nu}\phi\right]\,,\label{12}\\
{\cal F}_{\phi}[g_{\mu\nu},\phi]&=&\frac{\gamma'}{4\Lambda}(\nabla_{\mu}\nabla_{\nu}\phi n^{\mu}n^{\nu}-(\nabla^{2}\phi))K+\frac{\gamma'}{4\Lambda}(\nabla_{\mu}\nabla_{\nu}\phi)K^{\mu\nu}\,.\label{12.1}
\end{eqnarray}
Note that, from the Euler-Lagrange equation, ${\cal E}_{\phi}[g_{\mu\nu},\phi]={\cal F}_{\phi}[g_{\mu\nu},\phi]$.

\section{Q-profile within A BTZ black hole in Horndeski gravity}\label{v2}

In this section, we describe our BTZ black hole and construct the profile of the hypersurface Q taking into account the influence of  Horndeski gravity.

The three-dimensional metric of the BTZ black
hole is defined as the three-dimensional metric as  \cite{Banados:1992wn,Banados:1992gq}: 
\begin{eqnarray}
ds^{2}
=\frac{{{L}}^{2}}{r^{2}}\left(-{f}(r)d{{t}}^{2}+d{y}^{2}+\frac{dr^{2}}{{f}(r)}\right)\,.
\label{13}
\end{eqnarray}

A condition that deals with static configurations of black holes, which can be spherically symmetric for certain Galileons, was presented in Ref. \cite{Bravo-Gaete:2013dca} to discuss the no-hair theorem. The no-hair theorem requires that the square of
radial component of the conserved current vanishes identically without restricting the radial dependence
of the scalar field, which implies:
\begin{equation}
\alpha g_{rr}-\gamma G_{rr}=0\label{14}\,, 
\end{equation}

or, equivalently
$$ \alpha g_{rr} + \frac{\gamma'}{\Lambda}G_{rr}=0 \,.$$
\noindent 
From this condition we have ${\cal E}_{\phi}[g_{rr},\phi]=0$. 
These  solutions can be asymptotically dS or AdS for   $\alpha/\gamma<0$ or $\alpha/\gamma>0$, respectively, as discussed in Refs. \cite{Anabalon:2013oea,Santos:2020xox}. From now on, we will work with the asymptotically AdS case in three dimensions such that ${\Lambda} = -3 / {L}^2$, where $L$ is the AdS radius. Then, since $\Lambda <0$, the parameters $\gamma$ and $\gamma'=-\gamma\Lambda$ have the same sign.

Thus, we consider just $\phi=\phi(r)$ and define  $\phi{'}(r)\equiv {\psi}(r)$. It can be shown that the equations ${\cal E}_{\phi}[g_{rr},\phi]={\cal E}_{rr}[g_{rr},\phi]=0$ are satisfied and will be used to calculate the horizon function  ${f}(r)$ and $\psi(r)$, so that:
\begin{eqnarray}
{f}(r)&=&
\frac{\alpha}{\gamma'}-\left(\frac{r}{{r}_{h}}\right)^{2},\label{15}\\
{\psi}^{2}(r)&=&-\frac{6(\alpha-\gamma')}{\alpha\gamma' r^{2}{f}(r)}.\label{16}
\end{eqnarray}
Performing the transformations \cite{Brito:2019ose}: 
\begin{eqnarray} 
{f}(r) \to\frac{\alpha}{\gamma'}\,{f}(r),\,\, 
&& {r}_h^2  \to\frac{\gamma'}{\alpha}\, {r}_h^2
\\
{L} \to  \sqrt{\frac{\alpha}{\gamma'}}\, L\,, && 
{t} \to  \frac{\gamma'}{\alpha}\, t \\
{y} &\to & \sqrt{\frac{\gamma'}{\alpha}}\, y 
\end{eqnarray}
in the above equations, one finds that the metric \eqref{13} is invariant, while the horizon function $f(r)$ and $\psi(r)$ read
\begin{eqnarray}
f(r)&=&1-\left(\frac{r}{r_{h}}\right)^{2},
\label{frnovo}\\ 
{\psi}^{2}(r)&=&-\frac{6(\alpha-\gamma')}{\alpha^2 r^{2}{f}(r)} \,. \label{psinovo}
\end{eqnarray} 
The scalar field given by Eq. (\ref{psinovo}) should be real, and since $f(r)\ge 0$, we impose the constraint 
\begin{equation}
    {\alpha}\le {\gamma'}\,,
\end{equation}
with three possibilities: 
\begin{itemize}
    \item  $\alpha=\gamma'$;
    \item $\alpha>0$ and $\gamma'>0$; 
    \item $\alpha<0$ and $\gamma'<0$. 
\end{itemize}
The first choice implies $\psi(r)=\phi'(r)=0$. The second gives  $\alpha/\gamma'<1$, while the third is satisfied for  $\alpha/\gamma'>1$. The physical consequences of these choices will be associated with the solutions of  three-dimensional Horndeski gravity to be discussed below.

The Hawking temperature implied by the metric \eqref{13}, is given by %
\begin{equation}\label{hawk}
T_{H}= 
\dfrac{1}{4\pi} |f'(r_h)|
=\frac{1}{2\pi r_{h}}\,, 
\end{equation}
which is equal to the temperature of the dual BCFT theory $T_{BCFT}=T_{H}$. 

Now, in order to construct the Q boundary profile, one has that the induced metric on this surface given by 
\begin{eqnarray}
ds^{2}_{\rm ind}=\frac{L^{2}}{r^{2}}\left(-f(r)dt^{2}+\frac{g^{2}(r)dr^{2}}{f(r)}\right)\,, 
\end{eqnarray}
where $g^{2}(r)=1+{y'}^{2}(r)f(r)$ with $y{'}(r)=dy/dr$. Then, the normal vectors on Q can be represented by 
\begin{eqnarray}
n^{\mu}=\frac{r}{Lg(r)}\, \left(0,\, 1, \, -{f(r)y{'}(r)}\right)\,.\label{17}
\end{eqnarray}
Fulfilling the no-hair theorem, meaning 
${\cal F}_{\phi}[h_{rr},\phi]=0$, one can solve the Eq. \eqref{8}, so that 
\begin{eqnarray}
y{'}(r)&=&\frac{(\Sigma L)}{\sqrt{1-\dfrac{\gamma'\psi^{2}(r)}{4\Lambda}-(\Sigma L)^{2}\left(1-\left(\dfrac{r}{r_{h}}\right)^{2}\right)}}\,, 
\end{eqnarray}
\noindent with $\psi(r)$ given by Eq. \eqref{psinovo}, so that: 

\begin{eqnarray}
y{'}(r)&=&\frac{(\Sigma L)}{\sqrt{1-\dfrac{\xi'}{r^{2}
\Lambda
\left(1-\left(\dfrac{r}{r_{h}}\right)^{2}\right)}-(\Sigma L)^{2}\left(1-\left(\dfrac{r}{r_{h}}\right)^{2}\right)}}\,,
\label{19}
\end{eqnarray}
where $\xi'$ is defined as 
\begin{equation}
     \xi' 
     =-\frac{3}{2} \frac{ \gamma'}{\alpha} \left(1 -\frac{\gamma'}{\alpha} \right)
     \,. 
\end{equation}
If we choose $\alpha=\gamma'$, then $\xi'=0$. The second possibility is to take $\gamma'/\alpha>1$, so that the parameter $\xi'$ is positive and can be large. Third, for $\gamma'/\alpha<1$, $\xi'$  is negative and small. From now on, we consider this third case, except when explicitly mentioned. 
Besides, we can introduce $\Sigma L=\cos(\theta{'}) $ where $\theta{'}$ is the angle between the positive direction of the $y$ axis and the hypersurface Q.

\begin{figure}[!ht]
\vskip 1cm
\begin{center}
\includegraphics[scale=0.48]{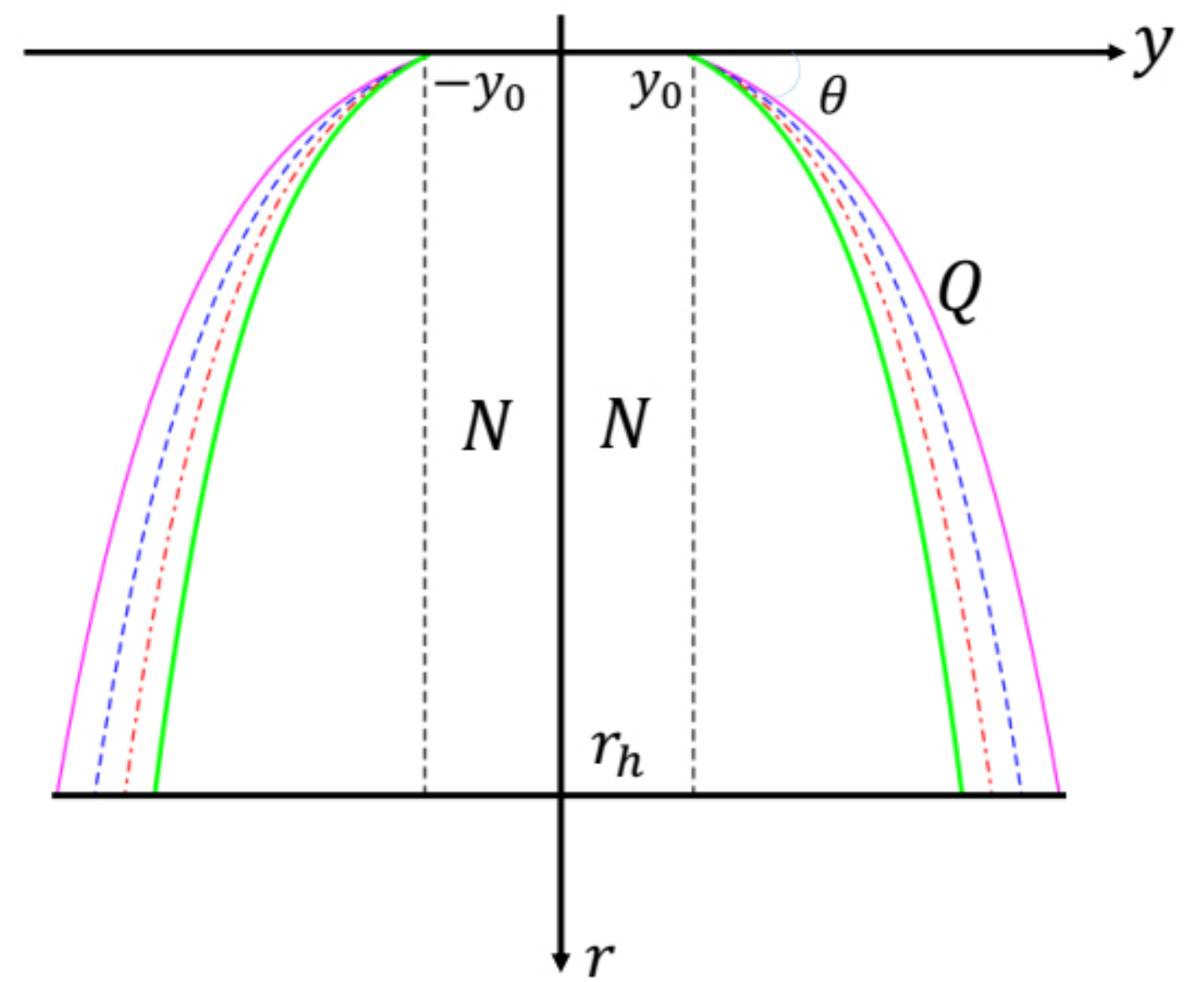}
\includegraphics[scale=0.48]{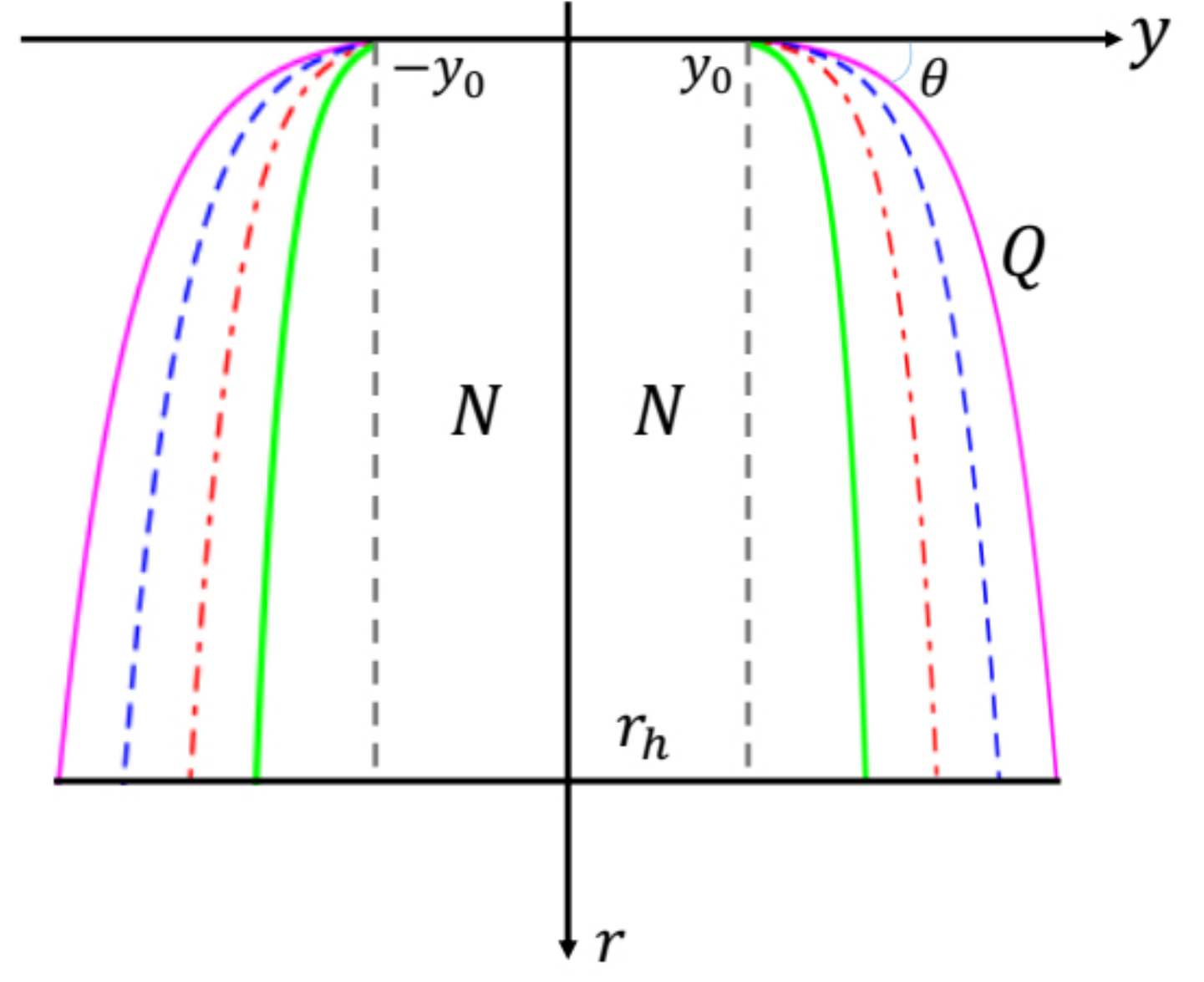}
\caption{Q boundary profile for the BTZ black hole within Horndeski gravity considering the values for $\theta'=2\pi/3$, $\theta=\pi-\theta'$,  $\alpha=-8/3$ with $\gamma'=0$ ({\sl solid}), $\gamma'=-0.1$ ({\sl dashed}), $\gamma'=-0.2$ ({\sl dot dashed}), and $\gamma'=-0.3$ ({\sl thick}). The  dashed parallel vertical lines represent the UV solution, Eq. \eqref{19.2}. The region between the negative and positive branches of curves Q represent the bulk N.
{\sl Left panel:}  we show the complete numerical solution  for the Eq. (\ref{19}). 
{\sl Right panel:}  we show the approximated solution for small values of $\xi'$, from the Eq. (\ref{19.3}). }\label{p0}
\label{ylinhaz}
\end{center}
\end{figure}

The equation for $ y{'}(r)$ can be solved numerically, and we can obtain the Q-profile for the $\gamma$-dependent Horndeski  terms (for $\xi'=0$ and $\xi'<0$) as shown in the left panel of Fig. \ref{p0}. 
Beyond the numerical solutions, we can analyze some particular cases regarding the study of the UV and IR regimes. Thus, for the UV case,  by performing an expansion at $r\to 0$ the Eq. (\ref{19})  becomes 
\begin{eqnarray}
y_{_{UV}}(r)=y_{0}+\frac{r\cos(\theta{'})}{\sqrt{-\xi'}}.\label{19.1}
\end{eqnarray}
In the above equation, considering for instance $\theta'=\pi/2$ or $3\pi/2$, we have
\begin{eqnarray}
y_{_{UV}}(r)=y_{0}={\rm constant}.\label{19.2}
\end{eqnarray}
This corresponds to a zero-tension limit $\Sigma\to 0$.

Now, for the IR case, we take  $r\to\infty$, so that Eq. \eqref{16} implies  $\psi(r\to\infty)=0$, and then $\phi=$ constant, which ensures a genuine vacuum solution. Plugging this result into Eq. (\ref{19}), in the limit $r\to\infty$,  we have
\begin{eqnarray}
y_{_{IR}}(r)=y_{0}+\ln(r/r_{h}).\label{19.22}
\end{eqnarray}

Another approximate analytical solution for $y(r)$ can be obtained by performing an expansion for very small $\xi'$ in Eq. (\ref{19}). Considering this expansion up to first order in $\xi'$, we obtain
\begin{eqnarray}
y_{_Q}\equiv y(r)&=&y_{0}+r_{h}\sinh^{-1}\left[\frac{r}{r_{h}}\cot(\theta{'})\right]
+\frac{\xi'L^{2}\cos(\theta{'})}{r_{h}}\tan^{-1}\left[\frac{r}{r_{h}\sqrt{1-\cos^{2}(\theta{'})f(r)}}\right]
\cr 
&+&\xi'L^{2}\cos(\theta{'})
\frac{\sqrt{1-\cos^{2}(\theta{'})f(r)}}{{r(-1+\cos^{2}(\theta{'}))^{2}}}
\left[{1+\frac{r^{2}\cos^{4}(\theta{'})}{r^{2}_{h}-r^{2}_{h}\cos^{2}(\theta{'})f(r)}}\right]
+\mathcal{O}(\xi'^{2})\,. \label{19.3}
\end{eqnarray}
It is worthwhile to mention that we consider $r_h^2 \gg \xi' L^2$, where $L$ is large to satisfy the AdS/CFT correspondence.
In the right panel of Fig. \ref{p0}, we plot the $y_{_Q}=y(r)$ profile from Eq. \eqref{19.3}, which represents our holographic description of BCFT within Horndeski's theory. Note that the bulk spacetime N is asymptotically AdS with two boundaries  M and Q. The interception of M and Q is represented by P in Fig. \ref{BCFT}. It is worthwhile to mention that the Q profile is obtained from the solution $y_{_Q}=y(r)$. 

Note that the UV solution  $y_{_{UV}}(r)=$ constant, Eq. \eqref{19.2} is similar to a lower-dimensional Randall-Sundrum (RS) brane, which is perpendicular to the boundary M.\footnote{A gravity theory containing solutions with nonzero tension of the RS branes was presented in Ref. \cite{Nozaki:2012qd}.} These RS-like branes could be represented in Fig \ref{p0} by the dashed parallel vertical lines. Further, as one increases the Horndeski parameter $\gamma'$, one can see that the surface Q gets closer to the RS-like branes.

\section{Holographic renormalization}\label{v3} 

In this section we present the holographic renormalization scheme in order to compute the Euclidean on-shell action which is related to the free energy of the corresponding  thermodynamic system.\footnote{One should notice that the free energy can also be calculated via the canonical thermodynamic potential by using the  black  hole  entropy  and  the  first  law  of  thermodynamics. This approach can be seen, for instance, in Ref. \cite{Gursoy:2017wzz}.}
 The holographic renormalization, as it is called within the AdS/CFT program,  is a steady approach to remove divergences  from infinite quantities on the gravitational side of the correspondence \cite{Henningson:1998gx,deBoer:1999tgo}. Such a renormalization on the gravity side will work similarly to the usual renormalization of the gauge field theory on the boundary.

Our holographic scheme takes into account the contributions of AdS/BCFT correspondence within Horndeski gravity. Let us start with the Euclidean action given by $I_{E}=I_{bulk}+2I_{bdry}$, i.e.,
\begin{eqnarray}
&&I_{bulk}=
-\frac{1}{16\pi G_{N}}\int_{N}{d^{3}x\sqrt{g}\left[(R-2\Lambda)-\frac{\gamma'}{2\Lambda}G_{\mu\nu}\nabla^{\mu}\phi\nabla^{\nu}\phi\right]}\cr 
&&-\frac{1}{8\pi G_{N}}\int_{M}{d^{2}x\sqrt{\bar{\gamma}}\left[(K^{(\bar{\gamma})}-\Sigma^{(\bar{\gamma})})+\frac{\gamma'}{4\Lambda}(\nabla_{\mu}\phi\nabla_{\nu}\phi n^{\mu}n^{\nu}-(\nabla\phi)^{2})K^{(\bar{\gamma})}\right.
 \left.+\frac{\gamma'}{4\Lambda}\nabla^{\mu}\phi\nabla^{\nu}\phi K^{(\bar{\gamma})}_{\mu\nu}\right]},\cr
 &&\label{BT}
\end{eqnarray}
where $g$ is the determinant of the metric $g_{\mu\nu}$ on the bulk N,   the induced metric and the surface tension on M are $\bar{\gamma}$ and $\Sigma^{(\bar{\gamma})}$, respectively and the trace of the extrinsic curvature on the surface M is $K^{(\bar{\gamma})}$.
 On the other hand, for the boundary, one has
\begin{eqnarray}
I_{bdry}&=&-\frac{1}{8\pi G_{N}}\int_{Q}{d^{2}x\sqrt{h}\left[(K-\Sigma)+\frac{\gamma'}{4\Lambda}(\nabla_{\mu}\phi\nabla_{\nu}\phi n^{\mu}n^{\nu}-(\nabla\phi)^{2})K+\frac{\gamma'}{4\Lambda}\nabla^{\mu}\phi\nabla^{\nu}\phi K_{\mu\nu}\right]}\nonumber\\
&&-\frac{1}{16\pi G_{N}}\int_{N}{d^{3}x\sqrt{g}\left[(R-2\Lambda)-\frac{\gamma'}{2\Lambda}G_{\mu\nu}\nabla^{\mu}\phi\nabla^{\nu}\phi\right]}.\label{BT1}
\end{eqnarray}
Through the AdS/CFT correspondence, we know that IR divergences in the gravity side correspond to the UV divergences at CFT boundary theory. This relation is known as the IR-UV connection. 

Thus, for the AdS-BTZ black hole, we can remove this IR divergence by introducing a cutoff $\epsilon$:
\begin{eqnarray}
I_{bulk}&=&\frac{1}{8\pi G_{N}}\int^{2\pi r_{h}}_{0}\int^{y}_{y_{0}}\int^{r_{h}}_{\epsilon}{\frac{L}{r^{3}}d\tau dydr}+\frac{1}{32\pi G_{N}}\int^{2\pi r_{h}}_{0}\int^{y}_{y_{0}}\int^{r_{h}}_{\epsilon}{\frac{L^{3}}{r^{3}}\frac{\gamma'}{\Lambda} G^{rr}\psi^{2}d\tau dydr}\nonumber\\
&&-\frac{1}{8\pi G_{N}}\int^{2\pi r_{h}}_{0}\int^{y}_{y_{0}}{\frac{L\sqrt{f(\epsilon)}}{\epsilon^{2}}d\tau dy}\,.\label{BT2}
\end{eqnarray}
Note that the coordinate $y$ in this equation,  associated with the AdS-BTZ black hole, is not the same as $y_{_Q}=y(r)$ related to the Q-profile discussed in section \ref{v2}. 
Then, we have for the bulk term:
\begin{eqnarray}
&&I_{bulk}=-\frac{L\Delta y}{8r_{h}G_{N}}\left(1-\frac{\xi'}{4}\right)+\mathcal{O}(\epsilon)\,, \label{BT3}
\end{eqnarray}
where $\Delta y\equiv y-y_0$. 

Analogously, for the boundary term we have
\begin{eqnarray}
I_{bdry}&=&\frac{1}{4\pi G_{N}}\int^{2\pi r_{h}}_{0}\int^{y_{_Q}}_{y_{0}}\int^{r_{h}}_{\epsilon}{\frac{L}{r^{3}}d\tau dy\, dr}\cr &+&\frac{1}{32\pi G_{N}}\int^{2\pi r_{h}}_{0}\int^{y_{_Q}}_{y_{0}}\int^{r_{h}}_{\epsilon}{\frac{L^{3}}{r^{3}}\gamma G^{rr}\psi^{2}d\tau dy\, dr}\label{BT4}\nonumber \\
&-&\frac{1}{8\pi G_{N}}\int^{2\pi r_{h}}_{0}\int^{r_{h}}_{\epsilon}{\frac{\Sigma L^{2}d\tau dr}{r^{2}\sqrt{1-(\Sigma L)^{2}f(r)}}}\cr
&+&\kappa\Sigma^{3}L^{2}\frac{\gamma'}{\alpha\Lambda}\left(1-\frac{\gamma'}{\alpha}\right)\frac{1}{8\pi G_{N}}\int^{2\pi r_{h}}_{0}\int^{r_{h}}_{\epsilon}{\frac{\Sigma L^{2}d\tau dr}{r^{2}\sqrt{1-(\Sigma L)^{2}f(r)}}}\,. 
\end{eqnarray}
This boundary action can be written as
\begin{eqnarray}
I_{bdry}&=&\frac{r_{h}L}{2G_{N}}\left(1-\frac{\xi'}{8}\right)\int^{r_{h}}_{\epsilon}{\frac{\Delta y_{_Q}(r)}{r^{3}}dr}\cr &+&\left(1-\frac{\xi'\cos^{3}(\theta{'})}{2}\right)\frac{L\cot(\theta{'})\csc(\theta{'})}{4G_{N}}+\mathcal{O}(\epsilon),\label{BT5}
\end{eqnarray}
where $\Delta y_{_Q}(r)\equiv y(r)-y_0$, with $y(r)$ given by Eq.(\ref{19.3}), and the Euclidean action $I_{E}=I_{bulk}+2I_{bdry}$ is given by:
\begin{eqnarray}
I_{E}&=&-\frac{L\Delta y}{8r_{h}G_{N}}\left(1-\frac{\xi'}{2}\right)+\frac{L}{G_{N}}\left(1-\frac{\xi'}{8}\right)w(\xi',r_{h})\nonumber\\
&+&\left(1-\frac{\xi'\cos^{3}(\theta{'})}{2}\right)\frac{L\cot(\theta{'})\csc(\theta{'})}{2G_{N}}\,,\label{BT6}
\end{eqnarray}
\noindent with
\begin{eqnarray}
&&w(\xi',r_{h})=\int^{r_{h}}_{\epsilon}{\frac{r_{h}\Delta y_{_Q}(r)}{r^{3}}dr}\,.\nonumber
\end{eqnarray}
Using the Q-profile for $\Delta y_{_Q}(r)$ from Eq. \eqref{19.3} in $w(\xi,r_{h})$, 
we can extract an approximated analytical expression for the Euclidean action $I_{E}$ as
\begin{eqnarray}\label{freeEBH}
I_{E}&=&-\frac{L\Delta y}{8r_{h}G_{N}}\left(1-\frac{\xi'}{4}\right)-\frac{L}{2G_{N}}\left(1-\frac{\xi'}{8}\right)\sinh^{-1}(\cot(\theta{'}))\nonumber\\
&+&\frac{\xi' q(\theta{'})L}{2G_{N}}+\frac{\xi' L^{3}h(\theta{'})\cot(\theta{'})}{2 G_{N}r^{2}_{h}}\label{BT6.1}\,,
\end{eqnarray}
\noindent where
\begin{eqnarray}
h(\theta{'})&=&-\frac{(1+\pi/2)}{2\sin(\theta{'})}+\frac{\cot^{3}(\theta{'})\cos^{2}(\theta{'})}{(1+\cos^{2}(\theta{'}))}\tanh^{-1}\left(\frac{\sqrt{2}\cos(\theta{'})}{\sqrt{1+\cos^{2}(\theta{'})}}\right)\nonumber\\
&-&\frac{(1+\cos^{2}(\theta{'})+3\cos^{4}(\theta{'})-3\cos^{6}(\theta{'}))}{3\sin^{5}(\theta{'})(1+\cos^{2}(\theta^{'}))}\,, 
\nonumber\\
q(\theta{'})&=&\left(\frac{1}{4}-\cos^{3}(\theta{'})\right)\cot(\theta{'})\csc(\theta{'})\,. \nonumber
\end{eqnarray}

Beyond the AdS-BTZ black hole, we can compute the Euclidean action for the thermal AdS solution considering $f(r)\to 1$. From  the equations (\ref{BT}) and  (\ref{BT1}), it is straightforward to get in this limit 
\begin{eqnarray}
I_{E}(0)=-\frac{L\Delta y}{8r_{h}G_{N}}\left(1-\frac{\xi'}{4}\right)\,.\label{BT6.2}
\end{eqnarray}

\section{BTZ black hole entropy in Horndeski gravity}\label{BTZentro}

In this section we compute the entropy related to the BTZ black hole considering the contributions of the AdS/BCFT correspondence within Horndeski gravity. From the free energy defined as
\begin{equation}\label{FE}
 \Omega=T_H\, I_E \,,  
\end{equation}
one can obtain the corresponding entropy as:
\begin{eqnarray}
S=-\frac{\partial\Omega}{\partial T_{H}}\,.\label{BT7}
\end{eqnarray}
By plugging the Euclidean on-shell action $I_E$, Eq. \eqref{freeEBH}, into the above equation, one gets
\begin{eqnarray}
S_{\rm total}&=&\frac{L\Delta y}{4r_{h}G_{N}}\left(1-\frac{\xi'}{4}\right)+\frac{L}{2G_{N}}\left(1-\frac{\xi'}{8}\right)\sinh^{-1}(\cot(\theta{'}))\cr&-&\frac{3\xi'L^{3}}{2r^{2}_{h}G_{N}} \cot(\theta{'}) h(\theta{'}) +\frac{\xi' L}{2G_{N}}q(\theta{'}).\label{BT8}
\end{eqnarray}
Recalling that the Hawking temperature, Eq. \eqref{hawk}, is a function of $r_h$, we should evaluate the profile from Eq. \eqref{19.3} at the horizon $r=r_h$. Then, one gets
\begin{eqnarray}\label{seno}
\frac{1}{2}\sinh^{-1}(\cot(\theta{'}))=\frac{\Delta y_{_Q}}{r_{h}}-\frac{\xi'L^{2}}{2r^{2}_{h}}\, b(\theta{'})
\end{eqnarray}
where 
\begin{eqnarray}\label{btheta}
b(\theta{'})=\cos(\theta{'})\tan^{-1}\left(\frac{1}{\sin(\theta{'})}\right)+\cot(\theta{'})\left(\frac{1+\cos^{2}(\theta{'})\cot^{2}(\theta{'})}{\sin^{2}(\theta{'})}\right)\,. 
\end{eqnarray}
Replacing Eq. \eqref{seno} in Eq. \eqref{BT8} one gets the total entropy with the bulk and boundary contributions with Horndeski terms: 
\begin{equation}
    S_{\rm total}= S_{\rm bulk + Horndeski} + S_{\rm boundary + Horndeski}\,, \label{St}
\end{equation}
where 
\begin{eqnarray}
S_{\rm bulk + Horndeski}&=&\frac{L\Delta y}{4r_{h}G_{N}}\left(1-\frac{\xi'}{4}\right) \\
 S_{\rm boundary + Horndeski}&=&\frac{L\Delta y_{_Q}}{r_{h}G_{N}}\left(1-\frac{\xi'}{8}\right)-\frac{\xi' b(\theta{'})L^{3}}{2r^{2}_{h}G_{N}}\left(1-\frac{\xi'}{4}\right)\nonumber \\&-&\frac{3\xi'L^{3} h(\theta{'})\cot(\theta^{'})}{2r^{2}_{h}G_{N}}+\frac{\xi' q(\theta{'})L}{2G_{N}}\,. 
\end{eqnarray}

One interpretation for this total entropy is to identify it with the Bekenstein-Hawking formula for the black hole: 
\begin{eqnarray}
S_{BH}=\frac{A}{4G_{N}}\label{BT9}\,.
\end{eqnarray}
Thus, in this case, from Eq. \eqref{St}, one has  
\begin{eqnarray}
A&=&\frac{L\Delta y}{r_{h}}\left(1-\frac{\xi'}{4}\right)+\frac{4L\Delta y_{_Q}}{r_{h}}\left(1-\frac{\xi'}{8}\right)-\frac{2\xi' b(\theta{'})L^{3}}{r^{2}_{h}}\left(1-\frac{\xi'}{8}\right)\nonumber \\&-&\frac{6\xi'L^{3} h(\theta{'})\cot(\theta^{'})}{r^{2}_{h}}+2\xi' q(\theta{'})L\,, \label{BT10}
\end{eqnarray}
where $A$ would be the total area of the AdS-BTZ black hole with Horndeski contribution terms for the bulk and the boundary Q. Since the information is bounded by the black hole area, the equation \eqref{BT10} suggests that the information storage increases with increasing $|\xi'|$, as long as $\xi'<0$. 

Note that the Bekenstein-Hawking equation \eqref{BT9} is a semiclassical result \cite{Das:2010su, Almheiri:2020cfm}. In this sense our total entropy ($S_{\rm total}$), Eq. \eqref{St}, can be interpreted as a correction to the original Bekenstein-Hawking formula:
\begin{equation}
S_{total} = S_{\rm Bekenstein-Hawking} + S_{\rm Horndeski\,\, contributions} \,.
\end{equation}
It is worthwhile to mention that corrections in the entropy were studied, for instance, in Refs. \cite{Hendi:2010xr, Solodukhin:2011gn, Bamba:2012rv, Feng:2015oea}. In particular, we found compatible results with the ones in Ref. \cite{Feng:2015oea}, where they considered Horndeski gravity in $n$-dimensional spacetime $(n\ge 4)$  within the Wald formalism or the regularized Euclidean action. 

Considering the boundary entropy for the AdS-BTZ black hole with Horndeski gravity, from Eq. \eqref{St}, one has:
\begin{eqnarray}
S_{bdry}=\frac{L\Delta y_{_Q}}{r_{h}G_{N}}\left(1-\frac{\xi'}{8}\right)-\frac{\xi' b(\theta{'})L^{3}}{2r^{2}_{h}G_{N}}\left(1-\frac{\xi'}{8}\right)-\frac{\xi'L^{3} h(\theta{'})\cot(\theta^{'})}{2r^{2}_{h}G_{N}}+\frac{\xi' q(\theta{'})L}{2G_{N}}, \;\;\quad \label{BT11}
\end{eqnarray}
which is identified with the entropy of the BCFT corrected by the Horndeski terms parametrized by $\xi'$. If we put $\xi'\to 0$ we recover the results presented in Refs. \cite{Takayanagi:2011zk, Fujita:2011fp}. In addition, still analyzing Eq. \eqref{BT11}, due to the effects of Horndeski gravity, there is a nonzero boundary entropy even if we consider the zero-temperature scenario, similar to an extreme black hole. This can be seen if one takes the limit $T\to 0$ ($r_h \to \infty$) in Eq. \eqref{BT11}; then, one gets what we call the residual boundary entropy
\begin{equation}
S_{bdry}^{res}=\frac{\xi' q(\theta{'})L}{2G_{N}}\,. \label{BT11ext}
\end{equation}
Note that, since the entropy should be non-negative, this zero-temperature limit is only meaningful if $q(\theta')<0$, once $\xi'<0$. 
In particular, considering our approximate analytical solution Eq. \eqref{19.3}, this will be fulfilled for small or large $\theta'$, $0< \theta'< \sqrt{6/13} $  or $ \pi / 2 < \theta' < \pi$, respectively. On the other side,  in the region $ \sqrt{6/13} < \theta' < \pi/2 $, one has $q(\theta')>0$, and then the limit $T\to 0$ cannot be reached. In this case there should be a minimum nonzero temperature corresponding to zero entropy.  

\section{Thermodynamic quantities and results}\label{v4}

The thermodynamics of black holes was established in Refs. \cite{Hawking:1971tu, Bardeen:1973gs, Bekenstein:1973ur}, and in this section we present our numerical results for the  thermodynamic observables, from a BTZ black hole. We take into account the contribution of the AdS/BCFT correspondence within Horndeski gravity. All of the thermodynamic observables are derived from the renormalized free energy.

Motivated by the thermodynamics of black
holes, the AdS/CFT and AdS/QCD were benefited, due to the possibility to construct effective gauge
theories at finite temperature opening a myriad of applications. In particular, the holographic study of charged black holes was presented in Refs. \cite{Chamblin:1999tk, Chamblin:1999hg}. These ideas were then applied to some high-energy phenomenology at finite temperature. For an incomplete list, see Refs. \cite{Kubiznak:2016qmn, Bravo-Gaete:2014haa, Zeng:2016aly, Gubser:2008ny, Gubser:2008yx, Li:2011hp, Cai:2012xh, He:2013qq, Zhao:2013oza, Li:2014hja, Li:2017ple, Rodrigues:2018pep, Chen:2018vty, Chen:2019rez, Rodrigues:2018chh, Arefeva:2020vae, Ballon-Bayona:2020xls, Arefeva:2020bjk, Caldeira:2020sot, Rodrigues:2020ndy, Caldeira:2020rir}.

\begin{figure}[!ht]
\begin{center}
\includegraphics[scale=0.55]{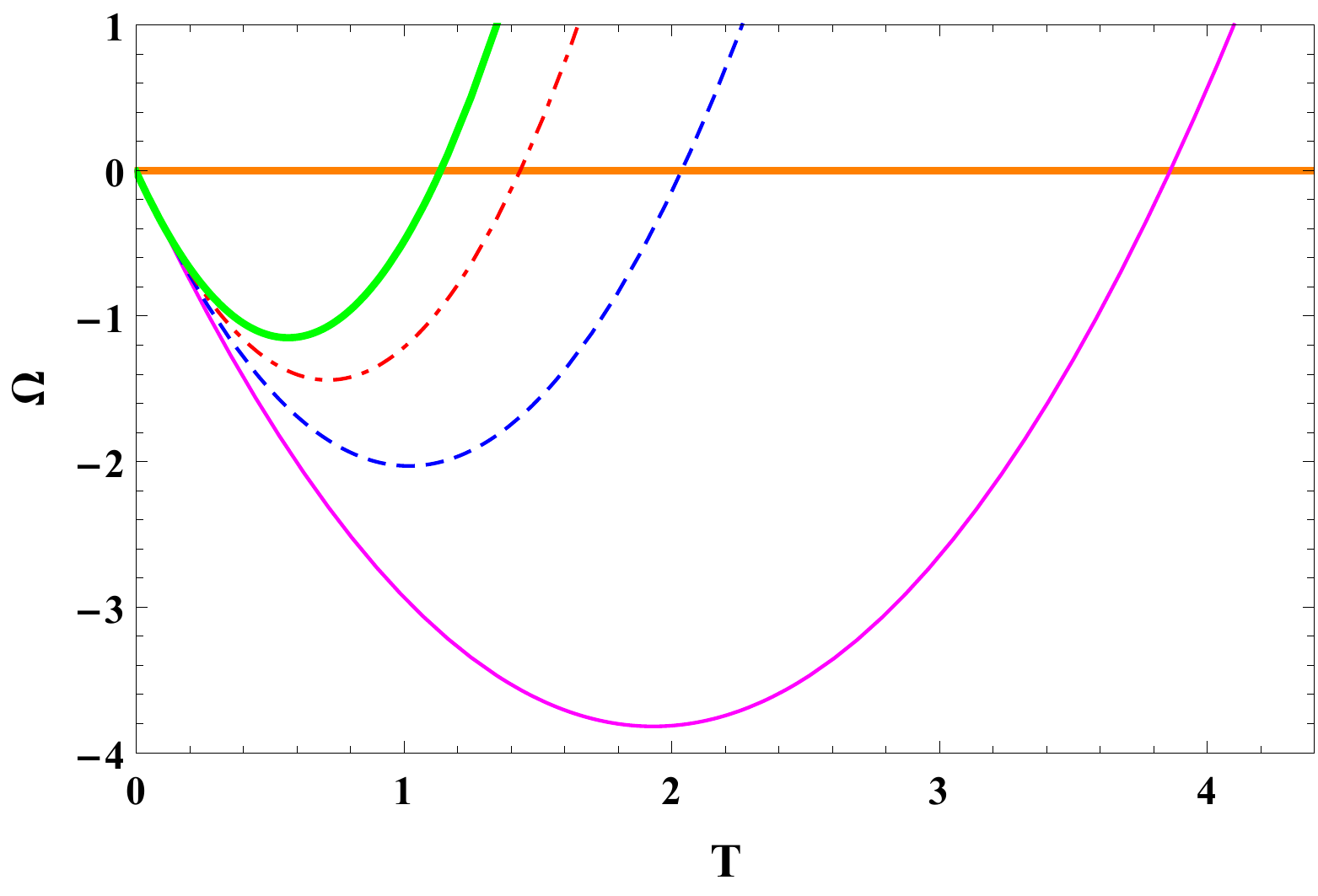}
\caption{Canonical potential or free energy as a function of the temperature and considering the influence of Horndeski gravity for the following values $\theta{'}=2\pi/3$, $\alpha=-8/3$,  with $\gamma'=-0.1$ (solid line), $\gamma'=-0.2$ (dashed line), $\gamma'=-0.3$ (dot dashed line), and $\gamma'=-0.4$ (thick line).}
\label{freeenergy}
\end{center}
\end{figure}

After this brief outlook, let us start our calculation from the differential form of the first law of thermodynamics, within the canonical ensemble. It can be written as:
\begin{equation}\label{1lei}
    d \Omega = -p dV - S dT\,,
\end{equation}
\noindent leading to 
\begin{equation}\label{omega}
    \Omega = \epsilon - TS \,,
\end{equation}
\noindent where $p$ is the pressure and  $\Omega$ is the canonical potential or free energy and $\Omega = T_{H}I_E $. The energy density is represented by $\epsilon$, $S$ is the entropy and $T$ is the temperature.
Besides for a fixed volume ($V \equiv 1$), one has:
\begin{equation}\label{1leientro}
     d \Omega = - S dT\,.
\end{equation}
Here, we present the behavior of the canonical potential or free energy, from Eq. \eqref{FE}. By analyzing Fig. \ref{freeenergy}, one can see that the canonical potential $\Omega$ has a minimum for each value of the Horndeski parameter $\gamma'$ which ensures a global condition of thermodynamic stability \cite{DeWolfe:2010he}. This picture also shows that there are critical temperatures where $\Omega =0$, depending on $\gamma'$. For $\Omega >0$ these solutions become unstable. The increase of the absolute value of $\gamma'$ induces a decrease of these critical temperatures.

\begin{figure}[!ht]
\vskip 1cm
\begin{center}
\includegraphics[scale=0.55]{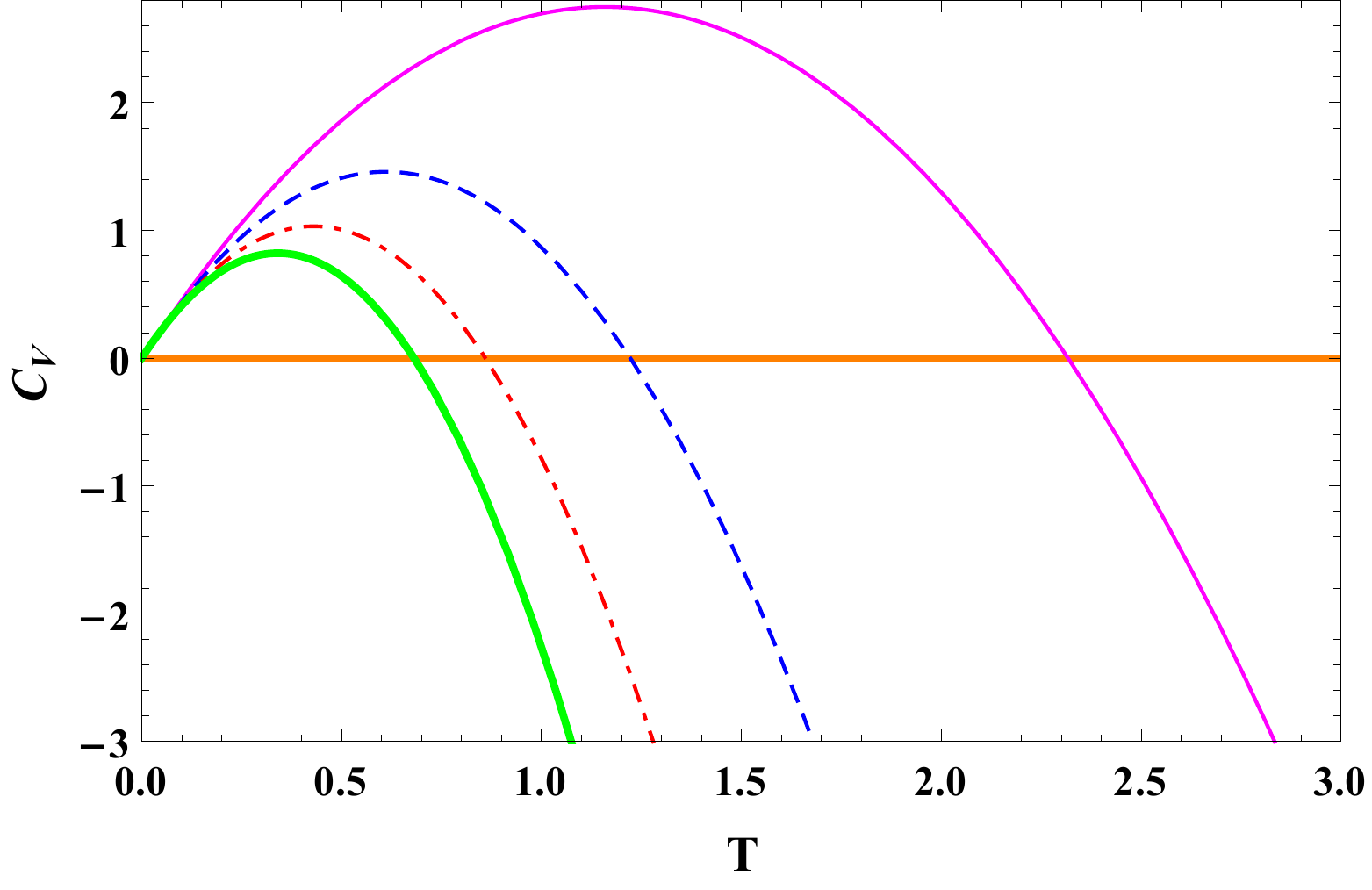}
\caption{Heat capacity as a function of the temperature and considering the influence of Horndeski gravity for the following values $\theta{'}=2\pi/3$, $\alpha=-8/3$,  with $\gamma'=-0.1$ (solid line), $\gamma'=-0.2$ (dashed line), $\gamma'=-0.3$ (dot dashed line), and $\gamma'=-0.4$ (thick line).}
\label{heatcapacity}
\end{center}
\end{figure}

\begin{figure}[!ht]
\vskip 1cm
\begin{center}
\includegraphics[scale=0.45]{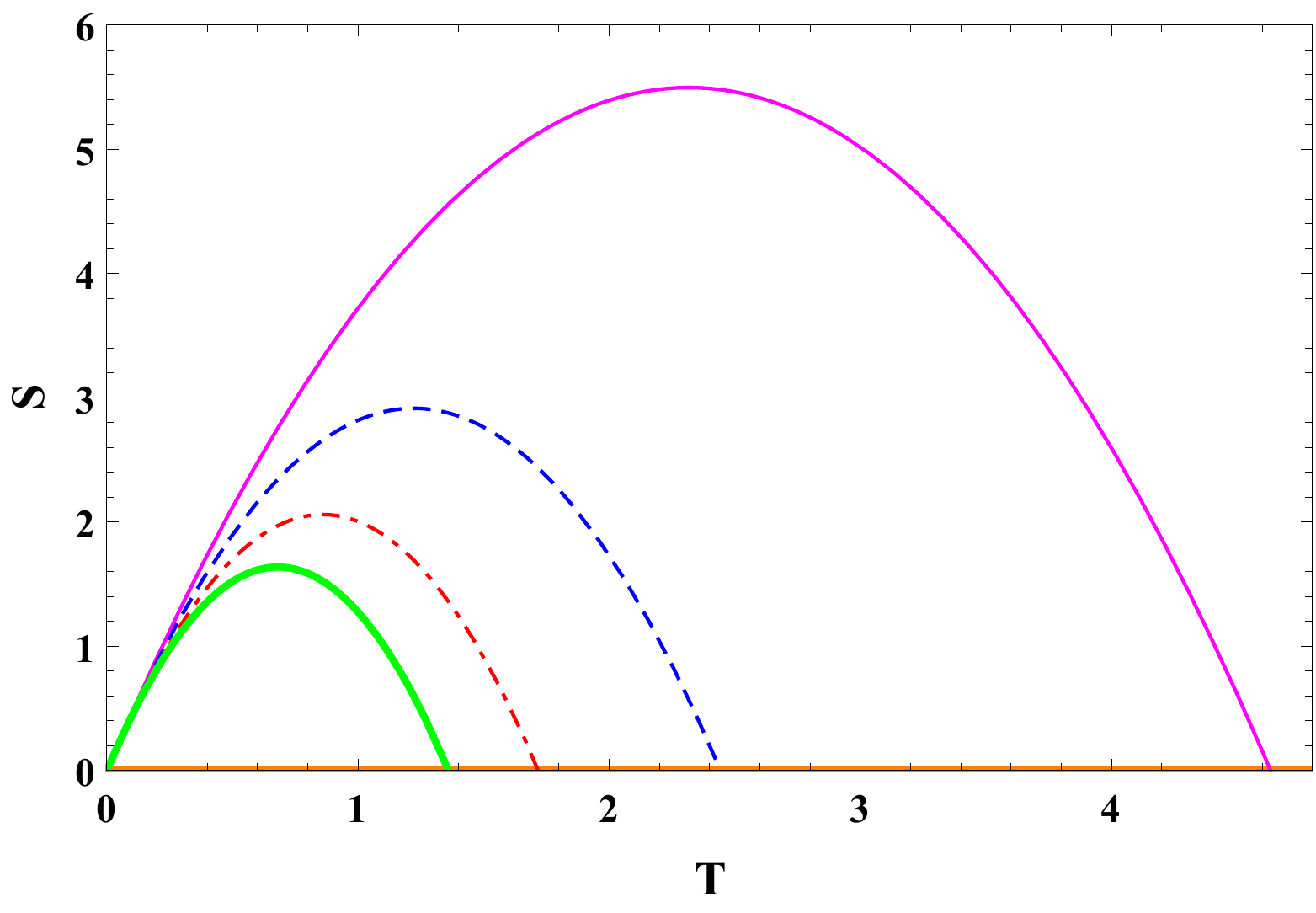}
\includegraphics[scale=0.45]{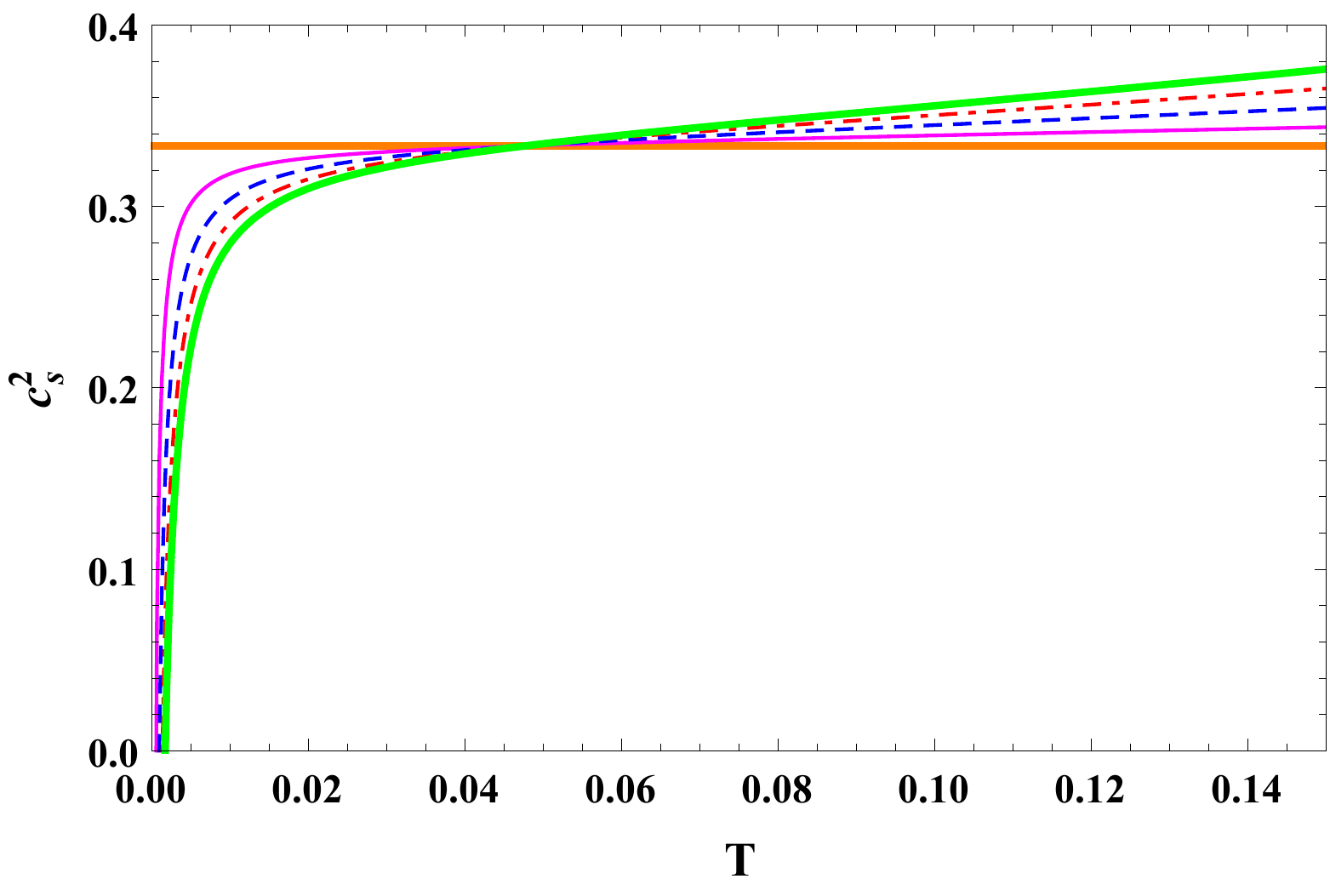}
\caption{ Entropy ({\sl left panel}) and sound speed ({\sl Right panel}) as a functions of the temperature and considering the influence of Horndeski gravity for the following values values $\theta{'}=2\pi/3$, $\alpha=-8/3$,  with $\gamma'=-0.1$ (solid line), $\gamma'=-0.2$ (dashed line), $\gamma'=-0.3$ (dot dashed line), and $\gamma'=-0.4$ (thick line).}
\label{entrovs}
\end{center}
\end{figure}

The next thermodynamic quantity that we analyze is the heat capacity $C_V$,  defined as: 
\begin{equation}
    C_V = T \left( \frac{\partial S}{\partial T}\right)_V = - T \left( \frac{\partial^2 \Omega}{\partial T^2} \right)\,.
\end{equation}
 In Refs. \cite{Ganai:2019lgc, Myung:2015pua, Ma:2013eaa, Hendi:2015wxa, Hendi:2016pvx} the authors discussed the positivity of the heat capacity and related it to the local black hole thermodynamic stability condition. This means that black holes will be thermodynamic stable if $C_V >0$. From Fig. \ref{heatcapacity} one can see that a black hole can switch between stable ($C_V>0$) and unstable ($C_V<0$) phases depending on the sign of the heat capacity. Also in Fig. \ref{heatcapacity} one can see the influence of Horndeski gravity on the temperature where the phase transition occurs.
The sound speed is defined as:
\begin{eqnarray}
    c_s^2 \equiv \frac{\partial p}{\partial \epsilon}
    = \frac{\partial T}{\partial \epsilon} \frac{\partial p}{\partial T} \,.
\end{eqnarray}
Identifying 
\begin{eqnarray}
    \frac{\partial T}{\partial \epsilon} = \left(\frac{\partial \epsilon}{\partial T}\right)^{-1} = C_V^{-1} \,;\qquad 
    \frac{\partial p}{\partial T} = S\,, 
\end{eqnarray}
one gets:\footnote{It is also very common to describe the sound speed as $c^2_s = \frac{\partial \ln{T}}{\partial \ln{S}}$.} 
\begin{eqnarray}
    c_s^2 &=& \frac{S}{C_V}\,.
\end{eqnarray}
 In Fig. \ref{entrovs}, we present the behavior of the entropy $S$  and sound speed  $c^2_s$ against the temperature achieved from our model. The entropy comes directly from  Eq. \eqref{BT8}. In the left panel one can see  the behavior of the entropy $S$ and the influence of Horndeski gravity. On the other hand, in the right panel, we show the sound speed and  the effects of Horndeski gravity which are more intense for $\gamma' = -0.4$. In this case it deviates from the value 1/3 associated with the conformal system. 

\begin{figure}[!ht]
\begin{center}
\includegraphics[scale=0.55]{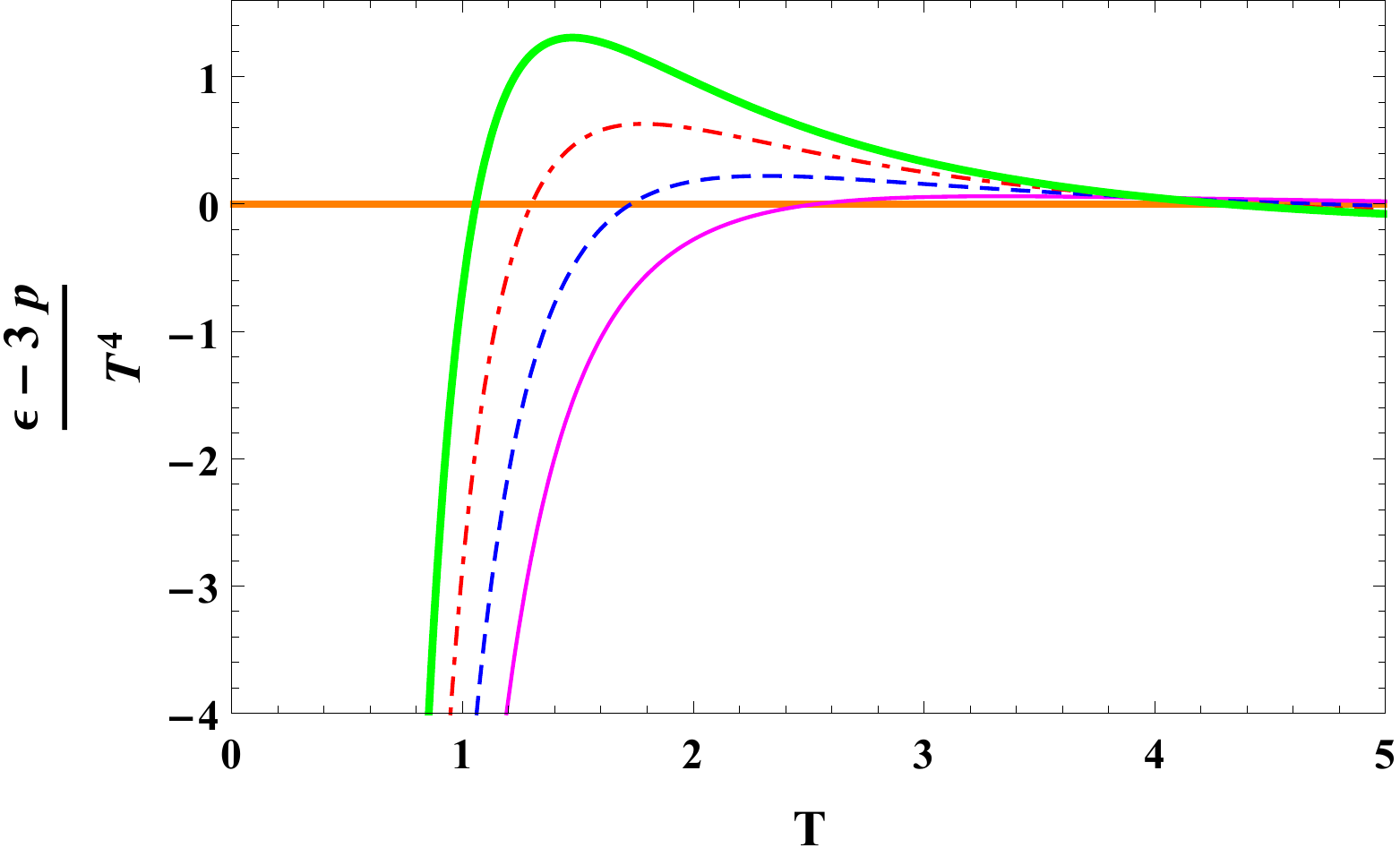}
\caption{Trace of the scaled energy-momentum tensor as a function of the temperature considering the following values $\theta{'}=2\pi/3$, $\alpha=-8/3$,  with $\gamma'=-0.1$ (solid line), $\gamma'=-0.2$ (dashed line), $\gamma'=-0.3$ (dot dashed line), and $\gamma'=-0.4$ (thick line).}
\label{trace}
\end{center}
\end{figure}

The last thermodynamic quantity that we present in this section is the trace of the energy-momentum tensor, defined as:
\begin{equation}
   \langle T^a_{\ \ a}\rangle = \epsilon - 3p = 4 \Omega + TS\,.
\end{equation}
In Fig. \ref{trace}, one can see the behavior of the scaled trace of the energy-momentum tensor ($\langle T^a_{\ \ a}\rangle/T^4$) as a function of temperature. It exhibits rather interesting behavior: for the small-temperature regime we have $\langle T^a_{\ \ a}\rangle \neq 0$., but  for the high-temperature regime, despite the influence of  Horndeski gravity, $\langle T^a_{\ \ a}\rangle \to 0$, which is an indication of a restoration of the conformal symmetry and therefore the emergence of a nontrivial BCFT.

\section{Hawking-Page phase transition}\label{v5}

In this section, we analyze the HPPT for a BTZ black hole considering the contributions of the AdS/BCFT correspondence within Horndeski gravity. The HPPT was originally proposed  in Ref. \cite{hawpage}, in the context of general relativity, where the authors discussed the stability and instability of  black holes in AdS space. The transition between the stable and unstable configurations characterizes a phase transition of first order with an associated critical temperature. 

In the context of the AdS/CFT program, the pioneering work in Ref. \cite{Witten:1998zw}, showed how to relate the temperature in gravitational theory with that associated with the gauge theory on the boundary.\footnote{Note that in Ref. \cite{Witten:1998zw} the Hawking temperature and Hawking-Page phase transition were associated with the temperature of deconfinement in QCD and the confinement/deconfinement phase transition. In this work we do not use such an interpretation.} For an incomplete list of works dealing with HPPT within the AdS/QCD context see, for instance, Refs. \cite{Cho:2002hq,Herzog:2006ra, Kajantie:2006hv, BallonBayona:2007vp, Rodrigues:2017cha, Rodrigues:2017iqi, Chen:2020ath, Li:2020khm,  Wang:2020pmb}. In particular, the HPPT within the BTZ black hole scenario was studied in e.g. Refs. \cite{Myung:2006sq, Eune:2013qs, Detournay:2015ysa, Myung:2015pua, Tang:2016vmu, Ganai:2019lgc}.\footnote{It is worthwhile to mention that only Ref. \cite{Eune:2013qs} used the holographic renormalization in order to compute the free energy. In all other listed references, the free energy was derived from the Bekenstein-Hawking entropy.}

The partition function for the AdS-black hole ($V_{E}$) is identified with minus the renormalized Euclidean action, Eq. \eqref{freeEBH}, $V_{E}=-I_E$, so that:
\begin{eqnarray}
V_{E}&=&\frac{L\Delta y}{8r_{h}G_{N}}\left(1-\frac{\xi'}{8}\right)+\frac{L \Delta y_{Q}}{2r_{h}G_{N}}\left(1-\frac{\xi'}{8}\right)\nonumber\\
&-&\frac{\xi' b(\theta^{'})L^{3}}{2r^{2}_{h}G_{N}}\left(1-\frac{\xi'}{8}\right)-\frac{\xi' q(\theta^{'})L}{2G_{N}}-\frac{\xi'L^{3} h(\theta^{'})\cot(\theta^{'})}{2r^{2}_{h}G_{N}}\,.\label{HP}
\end{eqnarray} 
\noindent Analogously, the partition function for the thermal AdS, is defined as $V_{E}(0)=-I_E(0)$, where $I_E(0)$ is given by Eq. \eqref{BT6.2}: 
\begin{eqnarray}
&&V_{E}(0)=\frac{L\Delta y}{8r_{h}G_{N}}\left(1-\frac{\xi'}{4}\right)\,.\label{HP1}
\end{eqnarray}
Now, we can compute $\Delta V_{E}$, so that:
\begin{eqnarray}
\Delta V_{E}=\frac{L\Delta y_{Q}}{r_{h}G_{N}}\left(1-\frac{\xi'}{8}\right)-\frac{\xi' b(\theta^{'})L^{3}}{2r^{2}_{h}G_{N}}\left(1-\frac{\xi'}{8}\right)-\frac{\xi'L^{3} h(\theta^{'})\cot(\theta^{'})}{2r^{2}_{h}G_{N}}-\frac{\xi' q(\theta^{'})L}{2G_{N}}\,.\label{HP2}
\end{eqnarray}
According to the HPPT prescription, the difference $\Delta V_E$ vanishes at the phase transition and $\Delta V_E < 0$ indicates the stability of the black hole. On the other hand, $\Delta V_E > 0$ points the stability of the thermal AdS space. 
\begin{figure}[!ht]
\vskip 1cm
\begin{center}
\includegraphics[scale=0.55]{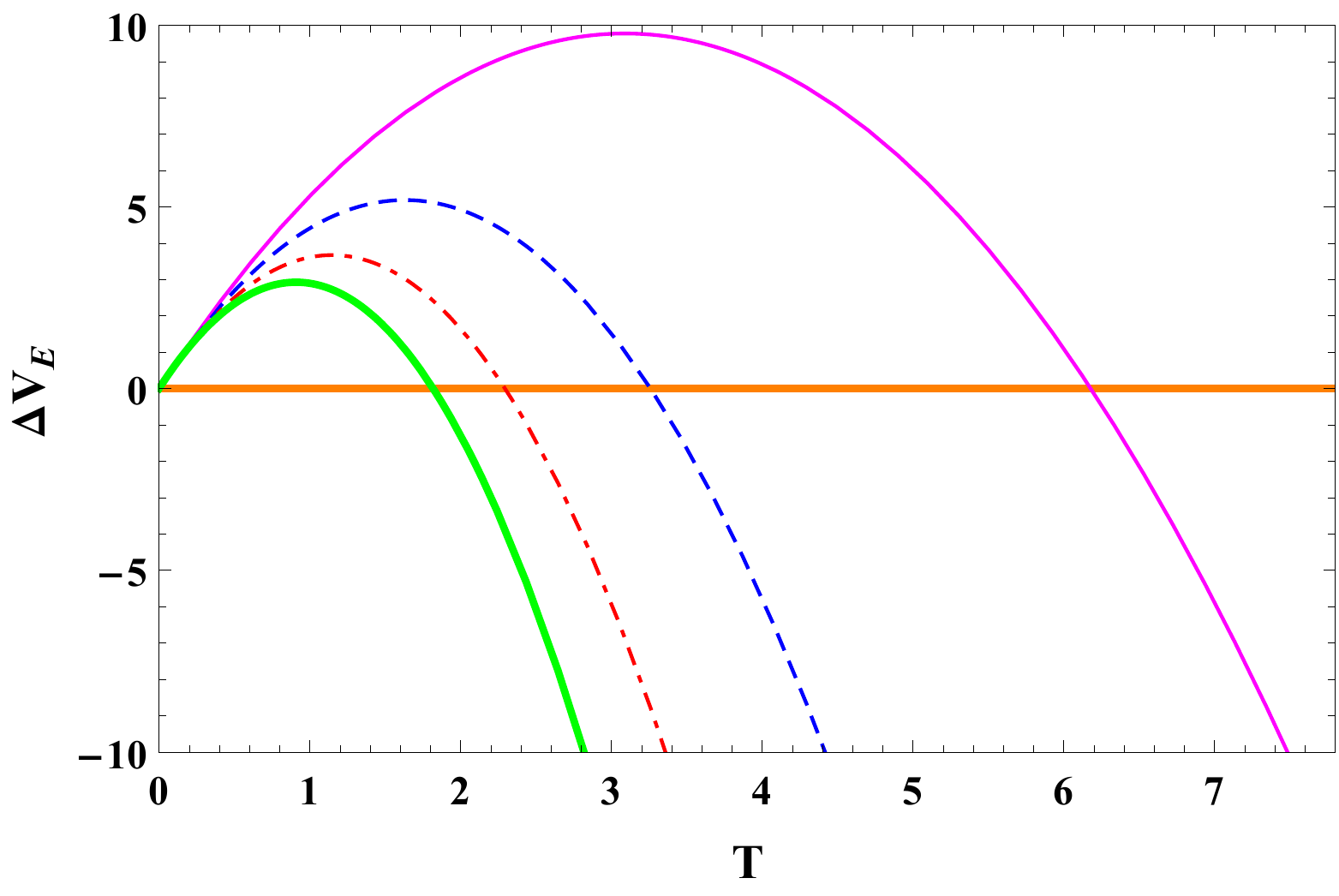}
\caption{Hawking-Page phase transition from Eq. \eqref{HP2} considering the following values $\theta{'}=2\pi/3$, $\alpha=-8/3$,  with $\gamma'=-0.1$ (solid line), $\gamma'=-0.2$ (dashed line), $\gamma'=-0.3$ (dot dashed line), and $\gamma'=-0.4$ (thick line). See the text for discussions.}\label{p01}
\label{planohwkhz}
\end{center}
\end{figure}

In Fig. \ref{planohwkhz}, we show the difference between the partition functions as a function of the temperature of the BTZ black hole in the AdS/BCFT correspondence and taking into account the contributions coming from Horndeski gravity. We see that the Horndeski effect decreases the HPPT critical temperature $T_c$, where $\Delta V_E=0$. Besides, the thermal AdS space is stable for low temperatures ($T<T_c$), while the AdS black hole is stable for the high-temperature regime ($T>T_c$). 

\section{Conclusion}\label{v6}

In this section we present our conclusions on the AdS/BCFT correspondence and BTZ black hole thermodynamics within Horndeski gravity. Considering the nonminimal coupling between the standard scalar term and the Einstein tensor we established our setup. Besides the three-dimensional bulk, we introduced a Gibbons-Hawking surface term and obtained the corresponding field equations. Then, using the no-hair-theorem, we found a consistent solution for the BTZ black hole. From this solution we constructed the Q profile on the two-dimensional boundary, which characterizes the AdS$_{3}$/BCFT$_{2}$ correspondence. In particular, we found an exact numerical solution and an approximate analytical one.\footnote{Note that these solutions for the boundary Q seem to describe a Randall-Sundrum brane in the limit of large Horndeski parameter.} 
These two solutions are shown in Fig. \ref{p0}, where one can see that the approximate solution describes qualitatively well the influence of the Horndeski term. So, starting in Sec. \ref{v3} and all subsequent sections, we only considered the approximated analytical solution. 
Using this solution, we performed a holographic renormalization procedure in order to get the Euclidean on-shell action for the thermal and the AdS-BTZ black holes. The identification of the Euclidean on-shell action with the free energy allowed us to compute the total entropy which is the sum of the contributions coming from the bulk and boundary both with Horndeski terms. From this total entropy and assuming the Bekenstein-Hawking formula we derived the corresponding total area for the AdS-BTZ black hole with Horndeski terms. We found that the total area grows with the absolute value of $\xi'$. This suggests that the information encoded on the black hole horizon also grows with $|\xi'|$. Another interpretation for the total entropy found in this work is that it represents a correction to the Bekenstein-Hawking formula. For the boundary entropy, it is remarkable that the influence of Horndeski gravity implies a nonzero or residual entropy in the zero-temperature limit $(r_h\to\infty)$, for a certain range of the angle $\theta'$. For another range of $\theta'$ 
the limit $T\to 0$ cannot be reached. In this case there it seems that there should be a minimum nonzero temperature corresponding to zero entropy.  

The free energy of the AdS-BTZ black hole with Horndeski gravity is depicted in Fig. \ref{freeenergy}. This picture shows the stability of these solutions for $\Omega <0$, up to a certain critical temperature depending on the Horndeski parameter $\xi'$.  
From this free energy we extracted the other relevant thermodynamic quantities, including the heat capacity, sound speed and trace anomaly. These results seem to be compatible with the ones expected from usual black hole thermodynamic properties. 

In Sec. \ref{v5}, we studied the Hawking-Page phase transition in the AdS/BCFT correspondence with Horndeski gravity. The modification coming from the Horndeski contribution allowed us to obtain this phase transition as a function of the temperature, as is usual in higher-dimensional contexts. This contrasts with the description of the HPPT given in Refs.  \cite{Takayanagi:2011zk, Fujita:2011fp}, where the authors plotted the free energy as a function of the Q profile tension.

Finally,  we  would  like  to  comment  that  these  theories  of  extended  gravity (such  as the  Horndeski  one)  and  beyond  Einstein’s  original  proposal where such theories take into account scalar fields and their couplings with gravity or accommodate higher-order terms in curvature invariants may provide new insights into aspects that can deepen our  knowledge  of  the  gravity  duals  of  conformal  field  theories  like the  AdS/BCFT correspondence.

\begin{acknowledgments}
We would like to thank Konstantinos Pallikaris, Vasilis Oikonomou, Adolfo Cisterna, and Diego M. Rodrigues for discussions. H.B.-F. is partially supported by Coordenação de Aperfeiçoamento de Pessoal de Nível Superior (CAPES) under finance code 001, and Conselho Nacional de Desenvolvimento Científico e Tecnológico (CNPq) under Grant No. 311079/2019-9.
\end{acknowledgments}
 

\end{document}